\documentclass[doublecol]{epl2} 
\usepackage{amssymb}
\newcommand{\M}[1]{\mathcal{#1}}
\newcommand{\h}[1]{\hat{#1}}
\newcommand{\tl}[1]{\tilde{#1}}

\newcommand{\ic}{\textmd{i}}
\newcommand{\avf}{\tilde{f}}

\title{On dissipation in crackling noise systems}

\author{Reinaldo Garc\'ia-Garc\'ia\inst{1,2}\thanks{E-mail: \email{reinaldomeister@gmail.com}}}
\institute{                    
  \inst{1} Laboratoire de Physico-Chimie Th\'eorique, CNRS UMR Gulliver 7083, PSL Research University,
ESPCI $-$ 10 rue de Vauquelin, 75231 Paris cedex 05, France\\
 \inst{2} PMMH, CNRS UMR 7636, PSL Research University, 
 ESPCI $-$ 10 rue de Vauquelin, 75231 Paris cedex 05, France
}

\abstract{
We consider the amount of energy dissipated during individual avalanches at the depinning transition of disordered and athermal elastic systems. Analytical
progress is possible in the case of the
Alessandro-Beatrice-Bertotti-Montorsi (ABBM) model for
Barkhausen noise, due to an exact mapping between the energy released in an avalanche and the area below a Brownian path until its first zero-crossing. 
Scaling arguments and examination of an extended mean-field
model with internal structure show that dissipation relates to a critical exponent recently found in a study of the rounding of the depinning transition in presence
of activated dynamics.
A new numerical method to compute the dynamic exponent at depinning in terms of blocked and marginally stable configurations is proposed, and a kind of `dissipative anomaly'\textendash
with potentially important consequences for nonequilibrium statistical mechanics\textendash is discussed. We conclude that
for depinning systems the size of an avalanche does not constitute by itself a univocal measure of the energy dissipated. 
}

\begin{document}

\maketitle

\section{Introduction}
Athermal disordered systems are ubiquitous in nature. When gently driven, they tend to evolve in a very intermittent manner, with quiescent periods
followed by activity burst, or avalanches, where the accumulated energy is released to the environment. For elastic manifolds driven in presence
of disorder, this occurs when the external force $f$ coincides with a critical force $f_c$ below which the manifold is stuck, and above which
it moves with a finite velocity. At $f=f_c$ the state of the system corresponds then to a dynamic depinning transition which is characterized by
such intermittent behavior, with power-law-distributed sizes and durations of individual avalanches.
Slightly above $f_c$ the proximity of the critical point is reflected by the power-law vanishing of the manifold
velocity, $v\sim (f-f_c)^\beta$, and the divergence of a correlation length, $\xi\sim(f-f_c)^{-\nu}$~\cite{KARDAR199885}.

The model of an elastic manifold in a disordered landscape is a good starting point for studying many systems exhibiting depinning-like behavior,
like contact lines~\cite{PhysRevLett.83.348,PhysRevB.53.3520}, domain walls in magnets~\cite{nattermann1998theory, ABBM-Classic, ABBM-Colaiori},
 charge density waves~\cite{RevModPhys.60.1129,PhysRevB.47.3530}, and vortex lines in superconductors~\cite{PhysRevB.52.1242}. The main focus
 of most studies is typically the critical force and dynamic observables, i.e., avalanche observables like sizes, durations, shapes and 
 velocities~\cite{NF-critical-CDW,Middleton1,Avalanche-velocity,Non-steady-ABBM,Avalanche-dynamics,Avalanche-shape}.
 Dissipation, however, has received far less attention in these systems even though it constitutes an important ingredient of their phenomenology.
 Some previous research has been devoted to the problem of dissipation at the depinning transition, even challenging the  typical assumption of pure viscous friction giving rise to the
 relaxational dynamics considered in the manifold model (see eq.~(\ref{dynamics}) below), e.g.,~\cite{Moulinet2004}, however, within the range of validity of linear friction, dissipation
 has not been systematically studied to the best of our knowledge.
 
 The main goal of this Letter is to give a first step in this direction. Our motivation
 is more than academical: the systematic study of dissipation may open new possibilities to better understand how to quantify more accurately 
 the energy released by earthquakes in terms of their extension, or to establish a link between the magnitude of plastic deformations and the structure 
 of local energy barriers in amorphous materials~\cite{PhysRevLett.102.235503,PhysRevB.81.064204,Yield-us}. 
 
Avalanche-size measurements in crystal plasticity by acoustic emission techniques~\cite{doi:10.1179/imtr.1980.25.1.41} rely on the equivalence
 between size and energy, which is to say that the size distribution, $p(S)\sim S^{-\tau}$, and the (dissipated) energy distribution,
 $P(Q)\sim Q^{-\varrho}$, exhibit the same power-law decay, $\tau=\varrho$. This has been shown to be the correct scenario in a study of crystal plasticity~\cite{SALMAN2012219}.
 One of the main results of this Letter is that this is not the case at depinning because the equivalence between size and energy is only possible under very special and singular
 conditions. In general, we will see that the link between size and energy implies an unexplored connection between dissipation at
zero temperature and the rounding of the depinning transition in presence of activated dynamics.

In other matters, our results lead us to propose a method to numerically access the dynamic exponent
at the depinning transition by only using general properties of blocked configurations, illustrating the fundamental role of energy surface topology on dynamics. Furthermore, we show that the so far
established dynamic mean-field theory is not consistent with a mean-field description of dissipation, suggesting that a modified scheme able to capture dissipation and kinetics
on equal footing is in order.

\section{Defining dissipation}
Being the central object of the present study, we start by appropriately introducing dissipation. The Hamiltonian of a $(d+1)$-dimensional 
elastic manifold
in presence of quenched disorder is written as $\M H=\M H_{el}+\M H_{dis}+\M H_{w}$, with the elastic part of the 
energy prescribed as $\M H_{el}=(c/2)\int_x [\partial_x u(x)]^2$, where $u$ represents the local height of the manifold and $c$ is the elastic coefficient\footnote{For simplicity we specialize here
on short-range elasticity without any lost in generality.}.
The disordered part reads $\M H_{dis}=\int_x V(u(x),x)$, where $V(u,x)$ is a Gaussian random potential with zero mean and correlator
$\overline{V(u,x)V(u',x')}=R(u-u')\delta^d(x-x')$. 
The system is confined by a global parabolic potential, $\M H_{w}=(m^2/2)\int_x [w-u(x)]^2$, and driven by slowly
and monotonically changing $w$. 

We consider simple relaxational dynamics, $\eta\partial_t u(x,t)=-(\delta/\delta u)\M H|_{u(x,t)}$:
\begin{equation}
 \label{dynamics}
\eta\partial_t u(x,t)=c\partial_x^2 u(x,t)+m^2[w(t)-u(x,t)]+F(u(x,t),x),
\end{equation}
where $\eta$ is the drag coefficient and the random force, $F(u,x)=-\partial_u V(u,x)$, has zero mean and covariance
$\overline{F(u,x)F(u',x')}=\Delta(u-u')\delta^d(x-x')$, with $\Delta(u)=-R''(u)$. 
Let us consider a time interval $[0,T]$. We introduce the work performed on the system during that time interval as is customary in
stochastic thermodynamics, i.e., as the energy change associated to the variation of the
parameters in the Hamiltonian~\cite{Jarzynski},
$W=\int_t(\partial_t\M H )=m^2\int_{x,t}\dot w[w-u]$, with $t\in[0,T]$. Multiplying eq.~(\ref{dynamics})
by $\partial_t u$, integrating over $x$ and $t$ and using an integration by parts in the spatial variables, we can write after very simple algebra:
\begin{equation}
 \label{energy-balance}
 \Delta\M H=W-\eta\int_{x,t}[\partial_t u(x,t)]^2,
\end{equation}
where $\Delta\M H=\M H[u(\bullet,T);w(T)]-\M H[u(\bullet,0);w(0)]$ corresponds to the total energy change during the process. 
By simple identification of
eq.~(\ref{energy-balance}) with first law of thermodynamics, $\Delta\M H=W-Q$, where $Q$ is the amount of energy released to the environment
in the form of heat, we can associate dissipation with the second term in the r.h.s. of eq.~(\ref{energy-balance})
\begin{equation}
 \label{dissipation}
 Q=\eta\int_{x,t}[\partial_t u(x,t)]^2\ge0.
\end{equation}

It becomes apparent that the amount of energy dissipated during the evolution can be simply related to the work done against
viscous forces. We also remark that at finite temperature eq.~(\ref{dissipation}) is corrected by adding 
a term proportional to the random noise
accounting for thermal fluctuations, which leads to the classical definition of stochastic heat commonly
used in stochastic thermodynamics~\cite{Seifert-Review}. In that case dissipation has no longer a definite sign at the stochastic level.
\section{Scaling analysis}
To gain some intuition on how the energy released in an avalanche is related to its size, we start by performing a scaling analysis. The results of this section are expected
to be general as far as the main assumptions behind the scaling arguments remain valid. In particular, the main result of this section, Eq.~(\ref{rho-general}), is expected to hold
for short range and long range elasticity as well, by using the corresponding exponents in each case.
\begin{figure}[t]
\onefigure[width=7.8cm,height=5.4cm]{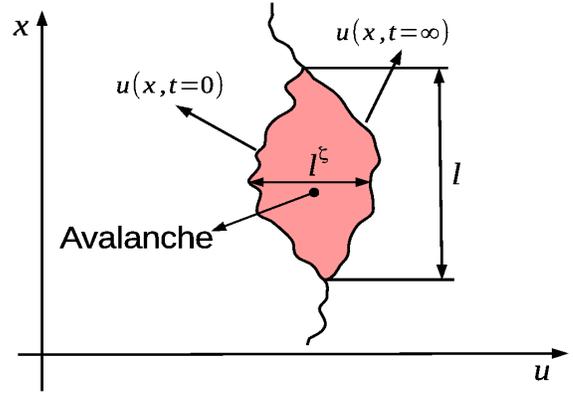}
\caption{Cartoon of an avalanche in $d=1$ with the corresponding length scales The size of the avalanche is the area of the shadowed region.}
\label{fig.1}
\end{figure}

In Fig.~\ref{fig.1} we schematically represent an avalanche for the case of a manifold in $d=1$. An initially stable configuration, $u(x,t=0)$, suffers from an instability
induced by the slow external driving, and part of the front evolves until the system finds a new globally stable configuration, $u(x,t=\infty)$. The
linear size of the portion of the manifold involved in the avalanche is called its extension and is denoted by $l$. The typical distance that the manifold
advances scales with the extension as $l^{\zeta}$, defining the roughness exponent of the interface at depinning, $\zeta$. The size of the avalanche corresponds
to the area of the shadowed region, i.e., the spanned `volume' of the avalanche, and can be expressed as $S=\int_{x,t}\dot u(x,t)~\sim l^{1+\zeta}$. In
$d$ dimensional isotropic media one has $S\sim l^{d+\zeta}$. Additionally, the typical duration of the avalanche, $T$, scales as $T\sim l^{z}$, defining the
dynamic exponent at the depinning transition, $z$. The `velocity' of the manifold then goes like $v\sim l^{\zeta-z}$, leading to the scaling
$Q=\eta\int_{x,t}[\partial_tu(x,t)]^2\sim l^{2\zeta+d-z}$ or equivalently, in terms of the size of the avalanche, $Q\sim S^{1-\psi_h}$, with
\begin{equation}
 \label{psi-h}
 \psi_h=\frac{z-\zeta}{d+\zeta}\equiv\frac{\beta}{(d+z)\nu-\beta},
\end{equation}
where in the second equality we have used the hyperscaling relation
$\beta=\nu(z-\zeta)$.
The notation $\psi_h$ is not accidental; the same exponent has been recently reported in the study of the thermal rounding of the depinning
transition~\cite{Rounding-Kolton}. More precisely, it determines the scaling of the velocity at $f=f_c$ with a small external
field $h$ providing a uniform activation rate for pinned portions of the manifold irrespective of the height of the local energy barrier to overcome,
$v\sim h^{\psi_h}$.
The precise meaning of this connection has remained so far elusive to us~\footnote{We have intensely discussed
about the origin of this connection with V. Lecomte. It was actually him who noticed that we were obtaining the same exponent of Ref.~\cite{Rounding-Kolton}.}. We will discuss about it elsewhere~\cite{R-V-D}.

The scaling relation between dissipation and size together with the asymptotic tail of the size distribution, $p(S)\sim S^{-\tau}$, determine the exponent of the power-law
decay of the dissipation distribution, $P(Q)\sim Q^{-(\tau-\psi_h)/(1-\psi_h)}$. We then find that scaling arguments predict a power law decay
$P(Q)\sim Q^{-\varrho}$, precising the exponent as
\begin{equation}
 \label{rho-general}
 \varrho=\frac{\tau-\psi_h}{1-\psi_h}.
\end{equation}

At this stage we can already note that the equality $\varrho=\tau$ only holds if $\psi_h=0$, in which case one also has $Q\sim S$. It turns out that this condition is actually very difficult
to meet in realistic systems, showing that at depinning the size of an avalanche does not scale linearly with the corresponding released energy. 

\section{Exact results: ABBM model}
\begin{figure}[t]
\onefigure[width=8.5cm,height=5.4cm]{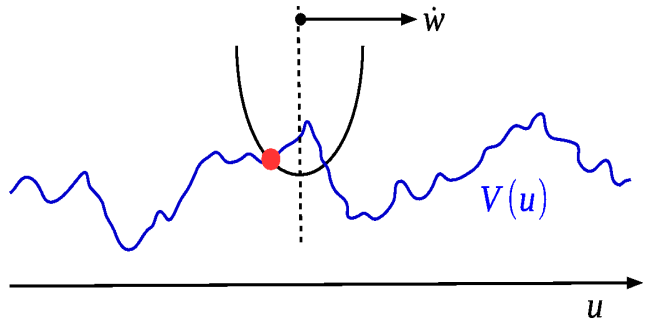}
\caption{Cartoon of a particle pulled by a spring moving at speed $\dot w\rightarrow0^+$
in presence of a random potential $V(u)$. The force deriving from this potential, $F(u)=-V'(u)$, describes a Brownian motion
in $u$.}
\label{fig.2}
\end{figure}

To go beyond scaling arguments, it is necesary to consider specific models.
The main goal of this paragraph is to analyze the asymptotic behavior of the probability density function (PDF) of the dissipation during an avalanche
in the simplest scenario.
With that purpose, we start by considering a $0-d$ manifold, i.e., a particle, pulled  by a spring in presence of a random potential, as schematically represented in 
Fig.~\ref{fig.2}.
The equation of motion for such a system reads
\begin{equation}
 \label{ABBM}
 \eta\partial_t u(t)= m^2[w(t)-u(t)]+F(u(t)).
\end{equation}

We specialize on the case of a Brownian force landscape, $\overline{F(u)\,F(u')}=2D\,\mbox{min}(u,u')$, where $D$ characterizes the strength of the disorder.
This choice corresponds to the classical ABBM model for domain wall dynamics and
Barkhausen noise in soft ferromagnets~\cite{ABBM-Classic,ABBM-Colaiori}. Under monotonous driving (we take $\dot w(t)=\mbox{const}\gtrsim0$) one can invoke Middleton
theorem~\cite{Middleton1,Middleton2,Middleton3} to ensure that if $\partial_t u(t=0)\ge0$ then $\partial_t u(t)\ge0\; \forall t$. The 
position of the particle is thus a monotonous function of time and, correspondingly, time can be reparametrized in terms of $u$. Taking
the time derivative of eq.~(\ref{ABBM}), and denoting by $v(u)$ the \emph{position-dependent} velocity, we have

\begin{equation}
 \label{reparametrization}
 \eta\partial_u v(u)=m^2\bigg[\frac{\dot w}{v(u)}-1\bigg]+\xi(u),
\end{equation}
where now $\xi(u)$ is a white noise in $u$, with zero mean and covariance $\overline{\xi(u)\xi(u')}=2D\delta(u-u')$. 
It is worth to make contact with the scaling analysis to predict what to expect for the ABBM model. In this case one has that the dynamic exponent
$z_{ABBM}=2$. As this problem is zero dimensional, the concept of roughness makes no sense. However, understanding the ABBM model as the zeroth-dimensional instance of the
Brownian Force Model (BFM) (see next section), for which the roughness exponent is $\zeta_{BFM}=4-d$ in $d\le4$, one can formally take the value $\zeta_{ABBM}=4$. With these two values and
the exponent of the size distribution, $\tau_{ABBM}=3/2$,
we readily
see from Eqs.~(\ref{psi-h}) and~(\ref{rho-general}) that we have $\psi_h=-1/2$ for the ABBM model, predicting the scalings $Q\sim S^{3/2}$ and $P(Q)\sim Q^{-4/3}$.

\begin{figure}[t]
\onefigure[width=8.0cm,height=5.4cm]{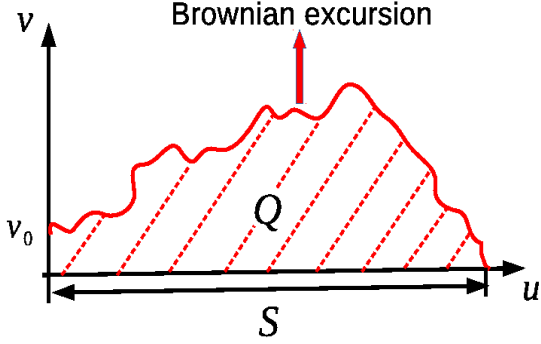}
\caption{Schematic representation of the size of an avalanche and its associated energy in the $v-u$ plane for the ABBM model.}
\label{fig.3}
\end{figure}

 Let us first consider the critical regime in which $m=0$. In this case
the velocity becomes a Wiener process in the position, $\eta\partial_u v(u)=\xi(u)$, and single avalanches can be identified as Brownian excursions of the velocity as a function
$u$. The size of an avalanche, $S$, which corresponds to the displacement of the particle until it stops, is nothing but the first-passage \emph{position} for $v(u)$,
i.e., $S$ corresponds  to the first value of $u$ for which $v(u)=0$. 
Dissipation, on the other hand, is proportional to the area below the Brownian path in the $v-u$, plane,
$Q=\eta\int_t [\partial_t u(t)]^2\equiv\eta\int_u v(u)$ (see Fig.~\ref{fig.3}).  The
PDF of the dissipated energy $Q$ during an avalanche then corresponds
to the distribution of the area until the first passage of a Wiener process, a well studied problem in mathematics
and physics, e.g.~\cite{darling1983,Louchard,Tacaks,Majumdar2005,Majumdar1,Dhar-Ramaswamy,Kearney}.
To the best of our knowledge, the calculation of the PDF of the energy released
during an avalanche in the ABBM model constitutes a new application for that theory.
With this analogy in mind, we can immediately write~\cite{Majumdar-Review}
 \begin{equation}
 \label{unconstrained-toy}
 P(Q) =\frac{1}{3^{2/3}\Gamma(1/3)}\frac{\eta v_0}{ D^{1/3} Q^{4/3}}\exp\bigg(-\frac{\eta^3v_0^3}{9D Q}\bigg).
 \end{equation}
 
 In eq.~(\ref{unconstrained-toy}), $v_0$ is a small cut-off initial velocity needed to regularize calculations
due to the singular nature of Wiener process. We point out that the existence of a small-scale cut-off is generic in physical systems,
however, the scale invariance of the underlying force landscape does not provide a natural small parameter, and the cut-off is set either by the driving
velocity or equivalently by an initial small velocity given to the particle after a small perturbation, which is the case we consider here.

The main message we extract from eq.~(\ref{unconstrained-toy}) is that at the depinning transition the distribution of the released
energy exhibits power law asymptotics for large $Q$, $P(Q)\sim Q^{-\varrho}$, modulated by some small-scale cut-off function, with specific value $\varrho=4/3$,
as we have already predicted above in terms of scaling arguments

To finish the present section, we briefly discuss the effect of a large-scale cut-off, wich appears when the mass is non-zero. For simplicity in the discussion,
we still consider that $\dot{w}=0$ and that the avalanche proceeds after exciting the particle with a small velocity $v_0$. It is known already that in that case it is impossible
to find a closed analytic expression for the distribution of the area below the Brownian path up to the first zero-crossing. However, one can at least extract the asymptotic behavior
of the distribution which decays for large $Q$ as follows~\cite{Majumdar1}
\begin{equation}
 \label{large-scale-cut-off}
 P(Q)\sim Q^{-3/4}\exp\bigg(-\sqrt{\frac{2 m^6 Q}{3 D^2}}\bigg).
\end{equation}
Two important comments are in order. First, note that the prefactor of the exponential has an exponent that does not correspond to the one predicted by scaling arguments.
This may imply that close to the large-scale cut-off there is a crossover in scaling. However, note that the large-scale cut-off, which is present in the exponential that dominates
the asymptotic behavior, is given by $Q_m\sim m^{-6}\propto S_m^{3/2}$, where $S_m=D/m^{4}$ is the known large-scale cut-off for the size of the avalanche. In other words,
the large-scale cut-offs for the size and dissipation PDFs \emph{do} scale as predicted by scaling arguments.
The relevance of the power-law prefactor close to the large-scale cut-off demands then further research.

\section{Dissipation in the Brownian Force Model}
We would like now to formally
test the predictions of scaling arguments by studying 
dissipation in the BFM~\cite{BFM-statics,Avalanche-velocity,Non-steady-ABBM,Avalanche-dynamics}.
The BFM corresponds to dynamics~(\ref{dynamics}) with a Brownian force in $u$,~$\overline{F(u,x)F(u',x')}=2D\,\mbox{min}(u,u')\,\delta^d(x-x')$.
It is worth noting that in comparison with the ABBM model discussed above, the BFM has internal structure and considers elasticity in an explicit
way.

The BFM is interesting in itself because it corresponds to the appropriate dynamic mean-field theory for a manifold
with internal dimension $d\ge 4$~\cite{Avalanche-velocity,Avalanche-dynamics}. It thus constitutes the starting point for studying
avalanche observables beyond mean-field at $d<4$ by using Functional Renormalization Group (FRG) in a loop expansion in $\epsilon=4-d$.
Velocity theory of the BFM is analytically tractable allowing a systematic study of many avalanche observables,
e.g,~\cite{Avalanche-velocity,Avalanche-shape,Local-size}.

A peculiarity of the BFM is that observables depending only on the position of the center of mass of the manifold exhibit, at any dimension $d$, the same 
statistics inferred from ABBM model.
In contrast, scaling analysis predicts the results for $Q$ depend on $d$. One can see that indeed this is the case by using known exponents for the BFM, $z_{BFM}=2$ and $\tau_{BFM}=3/2$
for all $d$, while $\zeta_{BFM}=4-d$ for $d\le 4$. One then gets from Eq.~(\ref{psi-h}) that $\psi_h=(d-2)/4$ 
leading to $Q\sim S^{(6-d)/4}$ and $P(Q)\sim Q^{-\varrho}$, with $\varrho=(8-d)/(6-d)$ from Eq.~(\ref{rho-general}). 
Notice that for $d=0$ we get $Q\sim S^{3/2}$ and $\varrho=4/3$, as previously obtained for the ABBM model.

To start, we note that by exploiting Middleton theorem one can derive an effective Markovian dynamics for the velocity $v(x,t)=\partial_tu(x,t)$~\cite{Non-steady-ABBM,Avalanche-dynamics}:
\begin{equation}
 \label{BFM-gen}
 \eta\partial_tv(x,t)=c\partial_x^2v(x,t)+m^2[\dot w-v(x,t)]+\sqrt{2D\,v(x,t)}\,\xi,
\end{equation}
with standard white noise of zero mean and covariance $\overline{\xi(x,t)\xi(x',t')}=\delta(t-t')\delta^d(x-x')$.
Let us first show that the large-scale cut-off for the dissipated energy scales with the one corresponding to the size of the avalanche in the correct manner, i.e., as predicted
by scaling analysis. This is simple to do if one introduces dimensionles variables, $v(x,t)=v_m\tl v(\tl x, \tl t)$, in Eq.~(\ref{BFM-gen}). One can readily show that~(\ref{BFM-gen})
becomes $\partial_{\tl t}\tl v=\partial^2_{\tl x}\tl v-\tl v+\dot{\tl w}+\sqrt{2\tl v}\,\tl \xi$, by choosing $v_m=(D/\eta c^{d/2})m^{d-2}$, and $\tl x=x/x_m$, $\tl t= t/t_m$, with
$x_m=\sqrt{c}/m$ and $t_m=\eta/m^2$. Then the large-scale cut-off for the size can be derived as $S_m=v_m x_m^d t_m= D/m^4$, while the large-scale cut-off for the dissipation reads
$Q_m=\eta v_m^2 x_m^d t_m= (D^2/c^{d/2})m^{d-6}$. It is now easy to check that $Q_m\propto S_m^{(6-d)/4}$ with an $m$-independent prefactor, which is the result we just discussed
above using generic scaling.

Having shown that the large-scale cut-offs exhibit the right scaling as they do in ABBM model, we now simplify the discussion by considering here the so called massless case in which
$m=0$. The initial state corresponds to a Middleton metastable state with vanishing interface velocity. An avalanche is then triggered by
giving a small force kick to the manifold at $t=0$. In this situation, Eq.~(\ref{BFM-gen}) is rewritten as
\begin{equation}
 \label{eff-velocity}
 \eta\partial_tv(x,t)=c\partial_x^2v(x,t)+\delta f(x)\delta(t)+\sqrt{2D\,v(x,t)}\,\xi(x,t).
\end{equation}
 Due to the lack of a large-scale cut-off in this setting for an infinite system, 
a uniform kick, $\delta f(x)=\delta f\; \forall x$, induces an avalanche of infinite size. One has then to impose a kick satisfying the integrability condition $\int_x \delta f(x)=\avf$, with
$0\lesssim\avf<\infty$. Furthermore, it is enough to consider a local kick, $\delta f(x)=\avf\delta^d(x)$; any other choice satisfying the integrability condition
gives the same tail for $P(Q)$, see the comment at the end of the present section.
The PDF of $Q$ is defined as $P(Q)=\overline{\delta(Q-\eta\int_{x,t}v(x,t)^2)}$.
Given that dissipation is positive definite, one can introduce its Laplace transform,
$G(\lambda)=\int_{Q\ge0}P(Q)\exp(-\lambda Q)\equiv\overline{\exp(-\lambda\,\eta\,\int_{x,t}v(x,t)^2)}$. 
In the Martin-Siggia-Rose path integral
representation of eq.~(\ref{eff-velocity}), we can write
\begin{equation}
 \label{Laplace-MSR}
 G(\lambda)=\frac{\int\M D[v,\h v]\exp\big(-I[v,\h v]-\lambda\,\eta\,\int_{x,t}v(x,t)^2\big)}{\int\M D[v,\h v]\exp\big(-I[v,\h v]\big)},
\end{equation}
with dynamic action
\begin{equation}
 \label{action}
I[v,\h v]=\int_{x,t} \ic\, \h v[\eta\partial_tv-c\partial_x^2 v-\avf\delta^d(x)\delta(t)]+D\int_{x,t}\h v^2\,v.
\end{equation}
Notice that by construction one
has $G(0)=1$, as demanded by normalization of $P(Q)$.
Taking the response field so that $\h v(x,t=\infty)=0$ we can, by
`integration by parts', rewrite the dynamical action as 
$
I[v,\h v]=-\ic\,\avf\,\h v(0,0)-\int_{x,t}\,\ic\, v(x,t)[\eta\partial_t\h v(x,t)+c\partial_x^2\h v(x,t)+\, \ic\,D\,\h v(x,t)^2].
$
We can now exactly perfom
a Gaussian integration
over $v$ in the numerator of eq.~(\ref{Laplace-MSR}) to write
\begin{equation}
 \label{Laplace-response}
 G(\lambda)=\M N\int \M D[\h v]\exp\big(-I_e[\h v]+\ic\,\avf\,\h v(0,0)\big).
\end{equation}

The integral is taken over all paths satisfying $\h v(x,t=\infty)=0$ and the normalization constant $\M N$ contains the denominator of~(\ref{Laplace-MSR}) and the determinant associated to
the Gaussian
integration, while the effective action reads 
\begin{equation}
 \label{eff-action}
 I_e[\h v]=\frac{1}{4\eta\lambda}\int_{x,t}[\eta\partial_t\h v+c\partial_x^2\h v+\,\ic\,D\,\h v^2]^2.
\end{equation}

A crucial step is to introduce dimensionless quantities as $\h v(x,t)= (c/D\ell_\lambda^2) h(x/\ell_\lambda,\, ct/\eta\ell_\lambda^2)$, with
$\ell_\lambda=(c^3/D^2\lambda)^{1/(6-d)}$. With this choice, all the terms inside the brackets in the effective action scale in the same way and can be factored out. Furthermore, the
multiplicative factor in front of the integral becomes $1/4$, while the second term inside the exponential in~(\ref{Laplace-response}) 
becomes $\ic \avf(c/D\ell_\lambda^2)h(0,0)$. All this allows to express the Laplace transform of $P(Q)$ as follows:
\begin{equation}
 \label{Laplace-final}
 G(\lambda)=\Phi\big((\lambda Q_f)^{\frac{2}{6-d}}\big),
\end{equation}
where $Q_f=[\avf^{(6-d)/2}/(c^d\, D^{(2-d)/2})]$ is a \emph{small-scale} dissipation cut-off (note that in $d=0$, under the identification $\avf=\eta v_0$, we obtain
the same typical scale as in eq.~(\ref{unconstrained-toy}), $Q_0=\eta^3 v_0^3/D$). Introducing the short-hand notation $h_0$ for $h(0,0)$, the function $\Phi(X)$, 
which satisfies $\Phi(0)=1$ by construction, is given as
\begin{equation}
 \label{Phi-funct}
 \Phi(X)\sim\int \M D[h]\exp\bigg(-\frac{1}{4}\int_{x,t}[\partial_t h+\partial_x^2 h+\,\ic\,h^2]^2+\ic\,X\,h_0\bigg).
\end{equation}

The tail of $P(Q)$ can now be derived as follows. First, write the inversion formula for the Laplace transform which, after the change of variable $\lambda\,Q=y$, can be expressed as $P(Q)=H(Q)/Q$, with 
\begin{equation}
\label{asympt-1}
 H(Q)=\frac{1}{2\pi\ic}\int_C dy\,\Phi\bigg(y^{\frac{2}{6-d}}\bigg[\frac{Q_f}{Q}\bigg]^{\frac{2}{6-d}}\bigg)\exp(y),
\end{equation}
for a suitably choosen contour $C$ parallel to the imaginary axis. Given that $\Phi(0)=1$, we see from~(\ref{asympt-1}) that $H(\infty)=\infty$. However, one can easily check
that $P(\infty)=0$ by using the final value theorem, $P(\infty)=\lim_{\lambda\to0}\lambda G(\lambda)$. Then, $\lim_{Q\to\infty}(H(Q)/Q)\equiv0$ exists,
which means that the asymptotic behavior of $P(Q)$ can be expressed using L'H\^{o}pital rule for large $Q$: $P(Q)\sim H'(Q)\sim Q^{-\varrho}$ to the leading order, with
\begin{equation}
 \label{confirmation}
 \varrho=\frac{2}{6-d}+1\equiv\frac{8-d}{6-d},
\end{equation}
confirming the prediction from scaling analysis. Before closing the present discussion it is worth showing that one indeed gets the same tail for $P(Q)$ under the action of any integrable force kick
satisfying $f(0)\neq0$ and $|x|^\alpha f(x)\to 0$ for $|x|\to\infty$ $\forall\;\alpha>0$.
In the case of an extended kick, we must substitute the term $\avf\,\h v(0,0)$ in eq.~(\ref{Laplace-response}) by $\int_x f(x)\,\h v(x,t=0)$, which under the same
rescaling as before behaves (ignoring unimportant prefactors) as $\ell_\lambda^{d-2}\int_x f(\ell_\lambda x)h(x,t=0)$. The key point is to note that $\ell_\lambda\to\infty$
at any $d\leq4$ for $\lambda\to0$, which is the sensitive limit to access the tail of the distribution. In that limit one has 
$$
\ell_\lambda^{d-2}\int_x f(\ell_\lambda x)h(x,t=0)\sim\avf\ell_\lambda^{-2}\,h(x=0,t=0)
$$ 
to the leading order,
which is equivalent to the local kick previously considered (more explicitly, $\ell_\lambda^d\, f(\ell_\lambda x)\to \avf\delta^d(x)$ for $\ell_\lambda\to\infty$).
\section{Discussion}
In the previous sections we have introduced the notion of dissipation during an avalanche, characterizing its behavior analytically in mean-field systems, 
and by means of scaling arguments in more general scenarios. The tail of the distribution of the energy dissipated during an avalanche corresponds to a power-law
decay with exponent $\varrho=(\tau-\psi_h)/(1-\psi_h)$ from scaling arguments, a prediction which is exactly satisfied by the BFM, in which case the exponent
reduces to the one reported in eq.~(\ref{confirmation}). This result also allows to correctly recover 
the $4/3$ law obtained analytically for the ABBM model.
We thus believe that our predictions indeed hold quite generally, although the results for the ABBM model suggest that one may
have a crossover in scaling that is not captured by the arguments provided previously. This demands a more detailed study of the dissipation in generic depinning systems.

Let us consider the necessary conditions to have dissipation and size of individual avalanches scaling in the same way. As previously commented, this would
demand the exponent $\psi_h$ to vanish, which, reading directly from eq.~(\ref{psi-h}), means that one should have $z=\zeta$. This situation is actually difficult to meet when considering realistic interfaces, however,
it is worth pointing out that it occurs in the BFM in two dimensions, as can be seen immediately by noting that $\zeta=4-d=2\equiv z$ in $d=2$.
Consistently, plugging $d=2$ in eq.~(\ref{confirmation}) gives $\varrho=\tau=3/2$. In any case, one sees that the relation $Q\sim S$ demands extremely fine tuning, and that in general
size and dissipation of individual avalanches at depinning scale differently. Furthermore, the scaling changes with dimension and other factors, implying that size does not provide a complete measure
for the dissipated energy. What this means precisely is that without additional information on the system, like knowing the dynamic exponent for instance, one cannot access dissipation by only 
meausring size.

We now propose a new method to determine the dynamic exponent at depinning. Note that one can in principle determine $z$ by measuring the distribution of avalanche
durations, but such an approach relies on simulating real-time dynamics, with a high computational cost. The method we propose allows to access
$z$ using the same quasistatic automata-like models commonly used to measure avalanche size distributions, which require far less computing resources.

Under quasistatic loading there is a clean separation between the time
scales of the driving and the relaxation. While energy accumulates until the manifold destabilizes, there is no dissipation, while once triggered, the avalanche
is so fast that during the discharge the external force has not enough time to do work.
The difference in energy between the activated state and the new Middleton state, which is a static quantity, is then equal, by conservation
of energy, to the heat released in the avalanche. One should then be able to determine the dynamic exponent $z$ from the knowledge of $\tau$, $\varrho$, and $d$ only; we proceed to show
that this is precisely the case

Starting from eq.~(\ref{psi-h}), we see that $z=(\zeta+d)(1+\psi_h)-d$. From the Narayan-Fisher (NF) prediction for $\tau$~\cite{NF-result-tau}, we write $\tau=2-2/(d+\zeta)$ (this relation has to be slightly
modified for long-range elasticity, but the modification is not essential for the upcoming analysis). We then get $d+\zeta=2/(2-\tau)$, which allows to write
$
z=[2(1+\psi_h)/(2-\tau)]-d.
$
Using now eq.~(\ref{rho-general}) we finally get
\begin{equation}
 \label{dynamic-exponent}
 z=\frac{2(2\varrho-\tau-1)}{(2-\tau)(\varrho-1)}-d,
\end{equation}
which is the desired result allowing to measure $z$. The method thus relies on modifying quasistatic simulations only in
the measurement step: In addition to initial and final position of the center of mass before and after the avalanches, one needs to compute the corresponding initial
and final energies of the system, dissipation being their difference. One can then build
the histograms for $S$ and $Q$ independently, accessing $\tau$ and $\varrho$, and correspondingly, $z$. We remark that, in view of the possibility
of a crossover in scaling, this method is expected to work well when sampling the central part of the distribution, far from the large-scale cut-off.

We now discuss on an important observation arising from our results. The dynamic tree-level theory (defined by the BFM) is called
precisely like that because it corresponds to the zeroth order in a loop expansion in powers of $\epsilon=4-d$ for dynamic observables in FRG. 
Correspondingly, the statistics of such observables do not depend on dimension $d$ at this level. We have however found
that dissipation exponent $\varrho$ contains corrections to \emph{all orders} in $\epsilon$ with respect to its
mean-field value at $d=4$, $\varrho=2$, a situation that we call `dissipative anomaly'. This clearly illustrates that dynamic mean-field theory
is not able to consistently describe dissipation within the same mean-field scheme.
It is then natural to expect those differences to survive beyond the tree-level. In our opinion, this situation is not exclusive
to the kind of systems we study here. On the contrary, we feel that it has a fundamental nature due to the way in which mean-field schemes deal with local fluctuations.
In any case, it is our hope that this observation will motivate the derivation of new mean-field approaches that consistently account for dissipation, which may prove relevant
when considering phase transitions out of equilibrium.

\section{Conclusions}
In this Letter we have analytically studied the statistics of the amount of energy dissipated during an avalanche 
at the depinning transition of an elastic manifold in a disordered landscape.
We have obtained the asymptotic form of the corresponding PDF, and our predictions were tested in a mean-field scenario.
Dissipation during an avalanche at criticality in the classical ABBM model can be exactly mapped
to the area under a Brownian path until its first zero-crossing. We have also proposed a numerical method
to determine the dynamic exponent at the depinning transition without simulating the full dynamics, i.e., by only using blocked and marginally stable configurations
in quasistatic simulations.

The simple scaling formula relating the dissipated energy and the size of an avalanche, $Q\sim S^{1-\psi_h}$, opens new
exciting research avenues due to the implied link between dissipation at depinning, and the rounding of the depinning transition
in presence of activated dynamics. We ignore so far the origin of this connection, and we expect that its understanding will motivate further research in the community.

Other possible extension of the present work
consists on studying the yielding transition in amorphous materials, in order to assess correlations between plastic (deformation) slips and local thresholds (related to local energy barriers).
On these lines, some parallels between depinning and amorphous plasticity 
may prove useful~\cite{Lin07102014,Damien-Botond}, although the breakdown of Middleton
theorem in the latter context poses a strong technical challenge.

Our identification of a dissipative anomaly where the dynamic tree-level theory contains dimensional corrections to all orders in $\epsilon=4-d$ for dissipation, while other avalanche observables
do not depend on dimension, seems to suggest that new mean-field schemes consistently accounting for dissipation are needed.
A possible route is to consider dynamic renormalization group calculations in ensembles which are constrained on dissipation.
We plan to pursue further results along this direction by studying simpler systems than those considered here. 

Finally, we believe that it is important to understand why in crystal
plasticity dissipation and size scale in the same way~\cite{SALMAN2012219}, while at depinning this is not in general the case, as discussed in this Letter.
Digging on that direction
may prove helpful to clarify the ultimate fundamental differences between depinning and plasticity.

\acknowledgments
This work was funded by CNRS and French National Research Agency under grant ANR-16-CE30-0022-03. We thank 
D. Vandembroucq, L. Truskinovsky and A. Hern\'andez Garc\'ia for enlightening discussions and critical suggestions. We also warmly thank V. Lecomte for revising the manuscript, for
pointing out the connection to Ref.~\cite{Rounding-Kolton}, and for the kind invitation to the
LIPhy at Universit\'e Grenoble-Alpes, where part of this work was completed.

\bibliographystyle{eplbib.bst}
\bibliography{./dissipation}

\end{document}